\documentclass[11pt]{article}
\usepackage{amssymb,amsmath}
\usepackage{graphicx}
\usepackage{epsfig}
\usepackage{color}

\input{tcilatex}

\begin{document}

\title{ Solving Solar Neutrino Puzzle via LMA MSW Conversion }
% \title{ Does Solar Neutrino Problem still exist?  }
% \textcolor[rgb]{0.98,0.00,0.00}{With 5 colorful} %
% \textcolor[rgb]{0.00,0.00,1.00}{areas marked}
%\thanks{% Supported by ....}}
\author{ 
{\large Q.Y. Liu$^{1,2}\footnote{email: qiuyu@ustc.edu.cn}$, B.L. Chen$^{1}$,
  J. Zhou$^{1}$, M.J. Luo$^{1}$, S.C. Jing$^{1}$ }
\bigskip
\\
{
$^{1}$ {\small \it Department of Modern Physics, University of Science
    and}
}
\\
{
{\small \it  Technology  of China, Hefei, Anhui 230026, China.}
}
\\
{
$^2$ {\small \it The Abdus Salam International Center for
Theoretical Physics,}
}
\\
{
{\small \it Strada Costiera 11,  34100, Trieste,
Italy.}
}
}

\date{}
\maketitle\vskip 12mm

\begin{abstract}

We analyze the existed solar neutrino experiment data and show the
allowed regions. The result from SNO's salt phase itself restricts
quite a lot the allowed region's area.  Reactor neutrinos play an
important role in determining oscillation parameters. KamLAND
gives decisive conclusion on the solution to the solar neutrino
puzzle, in particular, the spectral distortion in the 766.3 Ty
KamLAND data gives another new improvement in the constraint of
solar MSW-LMA solutions. We confirm that at 99.73$\%$ C.L. the
high-LMA solution is excluded.
\end{abstract}

\vspace{3mm}

\vskip 12mm

\vskip 5cm {\large \textbf{ %PACS: 12.15.Ji, ...
}} \vfill \eject
\baselineskip=0.36in \renewcommand{\theequation}{\arabic{section}.%
\arabic{equation}} \renewcommand{\thesection}{\Roman{section}}
\makeatletter
% '@' is now a normal "letter" for TeX
\@addtoreset{equation}{section} \makeatother
% '@' is restored as a "non-letter" character for TeX

\section{ Introduction }

 The electron neutrinos emitted from the sun disappear somewhere
when they travel to the earth. This is the famous solar
neutrino deficit, which is the almost forty years' ``Solar Neutrino
Problem''. There were many attempts to solve this
puzzle during the years. Some of them were tried to modify the solar model
in order to give a lower original neutrino flux, which conflict the
energy spectrum provided by
the 4 first-generation experiments: Homestake, Sage, Gallex and
Kamiokande \cite{CHLOR,SAGE,GALLEX,Kamiokande}.  Recent experiments have shown that the solar neutrino
oscillate by
$\nu _{e}\rightarrow \nu _{\mu ,\tau }$ inside the sun via MSW conversions.
This was proven by the Sudbury Neutrino Observatory (SNO) \cite{SNO}, and it
was confirmed by the reactor experiment KamLAND \cite{KamLAND}. The
former experiment detects $\nu _{e},$ $\nu _{e}+\nu _{\mu }+\nu _{\tau }$ and
$\nu _{e}+15\%(\nu _{\mu }+\nu _{\tau })$ three quantities on earth,
which correspond to CC, NC and ES interactions respectively;  KamLAND
observes $\overset{-}{\nu _{e}}\rightarrow $ $\overset{-}{\nu _{x}}$
neutrino oscillation channel.

\section{Solar neutrinos}

\bigskip The solar neutrino puzzle was solved by the neutrino oscillations $%
\nu _{e}\rightarrow \nu _{\mu ,\tau }$ inside the sun via MSW conversions.
This was proved by the Sudbury Neutrino Observatory (SNO) in Canada. And it
was confirmed by the laboratory base line experiment KamLAND in
Japan.

SNO is a 1000 ton heavy water Cerenkov detector mainly measuring
$^{8}B$ solar neutrinos. It consists of nearly 9450
photon-multiplier tubes and light concentrator units arrayed on a
geodesic support structure, with light water surrounding the
spherical acrylic vessel containing the $D_{2}O.$The first phase
of SNO data is from the pure $D_{2}O$. After that the
experimenters add up $NaCl$ (salt) to enhance the NC events rates.
This is called the second phase or "salt phase".

In analysis of the solar oscillation data \cite{Liu1}, we use
the $\chi^2$  defined as:

\begin{equation}
\chi _{\otimes }^{2}=\chi _{1gen}^{2}+\chi _{SK}^{2}+\chi _{SNO}^{2},
\label{Chi1}
\end{equation}
where $\chi _{1gen}^{2}$stands for Chlorine
and Gallium experiments. To calculate each individual chi square in the
right hand side of eq. (\ref{Chi1}), we use the so called
covariance approach:
\begin{equation}
\chi _{covar}^{2}=\sum_{n,m=1}^{N}(R_{n}^{\exp }-R_{n}^{theo})[\sigma
_{nm}^{2}]^{-1}(R_{m}^{\exp }-R_{m}^{theo})  .
\label{Chicovar}
\end{equation}
Here $R_{n}^{\exp }$and $R_{n}^{theo}$ correspond to experimental result and
theoretical value for the n-th data point. N=2,34,44 are for
 $\chi _{1gen\text{ }}^{2}$, $\chi _{SK}^{2}$, $\chi _{SNO}^{2}$
respectively. For getting a $R_{n}^{theo}$, the important step is to calculate the $%
\nu _{e}$ survival probability. We have used three methods to
check its consistency: the Parke formula \cite{Parke}; the
modified semi-analytic formula in \cite{Mont}; and the completely
numerical propagation. We found that the second way is the best,
considering both the calculation precision and the computer CPU
hour.

The covariant matrix of squared error $\sigma _{nm}^{2}$ can be written as

\begin{equation}
\sigma _{nm}^{2}=\delta _{nm}\mu _{n}\mu
_{m}+\sum_{k=1}^{K}c_{n}^{k}c_{m}^{k}.
\label{sigmamn}
\end{equation}

$\mu _{n}$ is the uncorrelated error of the n-th detected quantity for both
the experiment and the theory (such as the statistic uncertainty, or other
uncertainties which affect only one detectable value), and $c_{n}^{k}$ is
the correlated systematic error caused by the k-th correlated original
error (the original error may be the spectrum uncertainty, or the energy
resolution uncertainty, etc.). For a detectable value $R_{n}^{\exp }$, we
can say that its uncertainty $(R_{n}^{\exp }-$ $R_{n}^{theo})$ is in
the range of
$\pm \mu _{n}\pm c_{n}^{1}\pm c_{n}^{2}\pm ...\pm c_{n}^{k}.$

For $\chi _{1gen}^{2},$ the correlated errors are the 12 SSM uncertainties $%
X_{i}$ (the cross section factors
$S_{11},S_{13},S_{34},S_{1,14},S_{17}$; the $^{7}Be$ capture cross
section $C_{Be}$; the solar luminosity; metallicity Z/X; age;
opacity; the element diffusion and $S_{hep}$), whose relative
uncertainties $\Delta \ln X_{k}$ determine the correlated
uncertainties of the neutrino fluxes $\Phi _{i}^{SSM}$ through the
logarithmic derivatives

\begin{equation}
\alpha _{i,k}=\frac{\partial \ln \Phi _{i}^{SSM}}{\partial \ln X_{k}}
\label{alphaik}
\end{equation}

With this uncertainties and the matrix given in (\ref{sigmamn})
(\ref{alphaik}),
we can calculate the fractional uncertainties of the SSM neutrino fluxes:

\begin{equation*}
\Delta \ln \Phi _{i}^{SSM}=\sqrt{\sum_{k=1}^{12}(\alpha _{i,k}\Delta \ln
X_{k})^{2}}
\end{equation*}

For $\chi _{SNO}^{2}$ and $\chi _{SK}^{2}$, the only effective SSM
uncertainty is the $^{8}B$ neutrino flux uncertainty since the $hep$ neutrino
flux is too small. So, we uses free $^{8}B$
flux method \cite{Hol}, i.e., we float the flux near the central value
given in paper \cite{BP}
within $1\sigma $ uncertainty range, then find out which flux gives the
minimal $\chi ^{2}$. The $\chi ^{2}$ formula changes a little as

\begin{equation*}
\chi ^{2}=\sum_{n,m=1}^{N}(R_{n}^{\exp }-f_{B}\cdot R_{n}^{theo})[\sigma
_{nm}^{2}]^{-1}(R_{m}^{\exp }-f_{B}\cdot R_{m}^{theo})+\frac{(1-f_{B})^{2}}{
\sigma ^{2}}.
\end{equation*}

Thus to describe $\sigma _{nm}^{2}$ in these two experiments, we don't need to
consider the SSM uncertainties. The remaining original uncertainties that
can affect $\chi _{SK}^{2}$ are the $^{8}B$ spectrum shape error, the energy
scale uncertainty, the resolution uncertainty and an overall SK systematic
offset uncertainty. For SNO, the remaining original uncertainties are $^{8}B$
spectrum shape error, the energy scale uncertainty, the resolution
uncertainty, the vertex reconstruction uncertainty, the cross section
uncertainty, the neutron capture uncertainty, the neutron background
uncertainty, and low energy background uncertainty \cite{SNO}.

\begin{figure}[t]
\mbox{\epsfig{figure=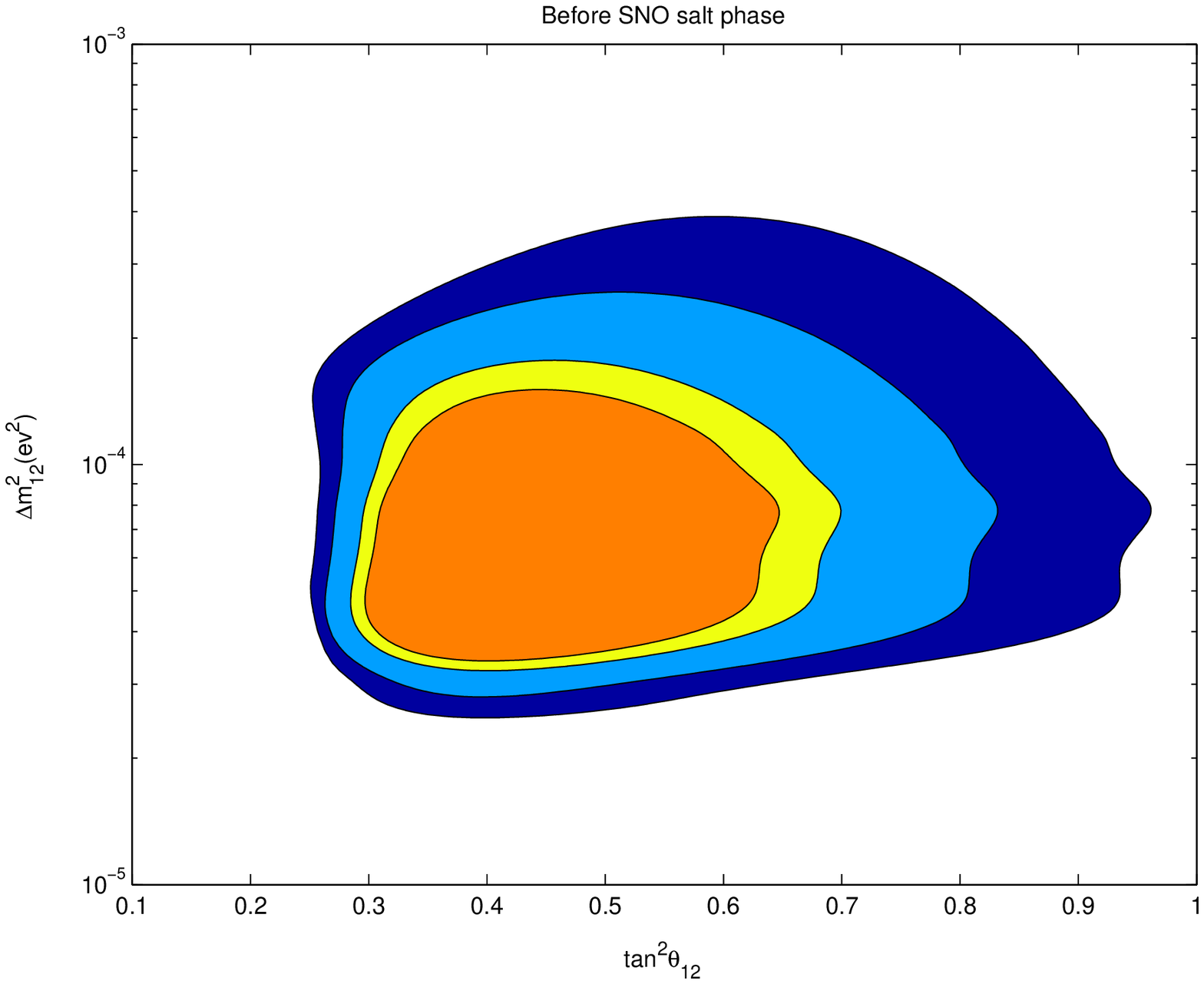,height=6cm}} \mbox{%
\epsfig{figure=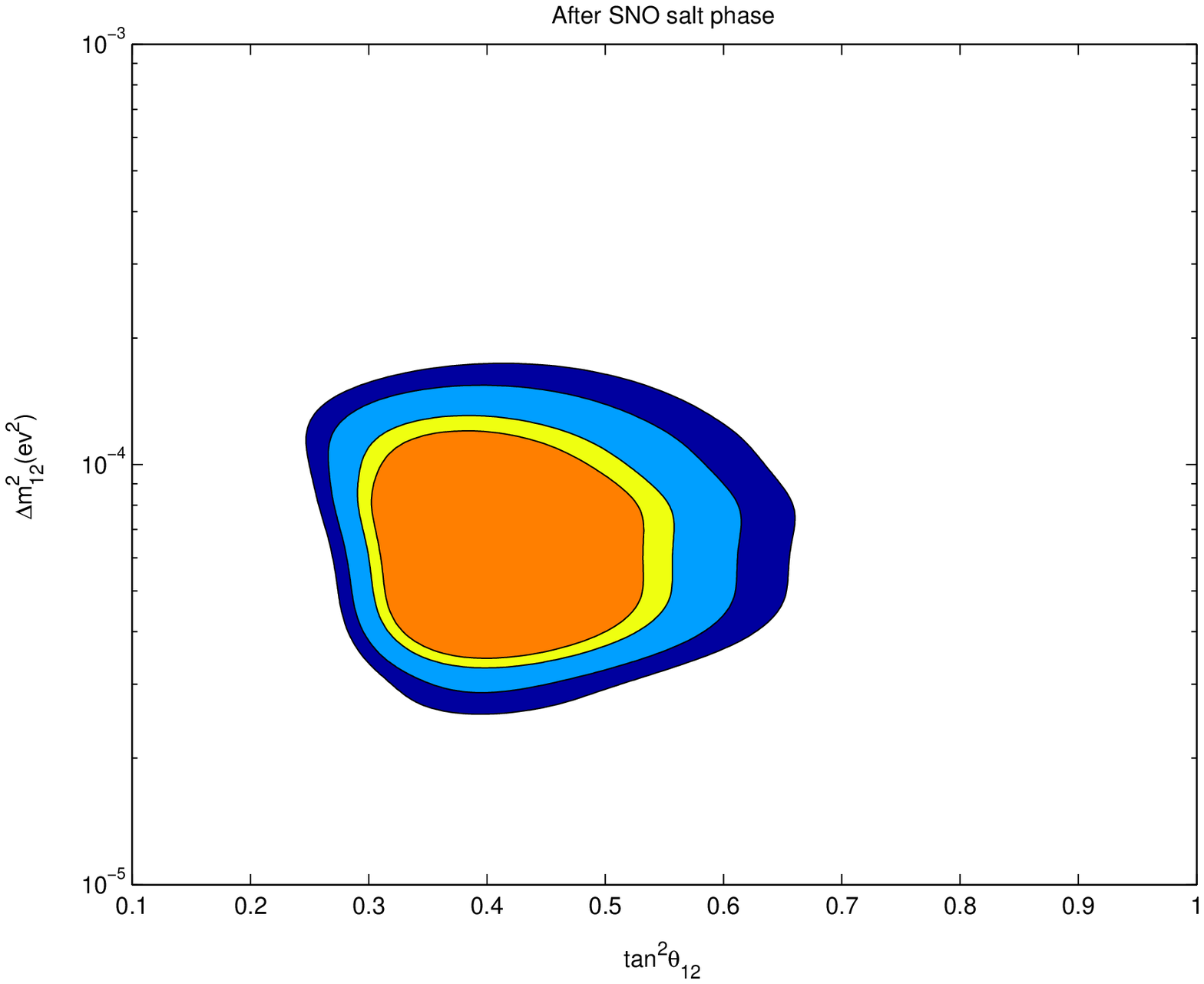,height=6cm}}
\caption{ The 90$\%$, 95$\%$, 99$\%$, and 99.73$\%$ C.L. allowed regions in
the $\Delta m_{12}^{2}-tan^{2}\protect\theta_{12}$ plane after SNO
experiments. Left corresponds to the result of SNO's first phase, the right
is the result of SNO's salt phase case. }
\label{fig:SNO}
\end{figure}

\bigskip Using this method we scan the solar neutrino $\tan ^{2}\theta
_{12}-\Delta m_{12}^{2}$\ parameter space, neglecting the tiny
correction from $\theta _{13}$ and$\ \Delta m_{13}^{2}.$ The allowed regions
are shown in figs. \ref{fig:SNO}.

\section{Manufactured neutrinos}

At present the only detected laboratory neutrinos which are sensitive to
$\Delta m_{12}^{2}-\tan\theta _{12}$ parameters are reactor neutrinos. KamLAND
is a first experiment to deal with these solar neutrino parameters completely from
manufactured source instead of from the sun. This reactor neutrino experiment
convinces scientists that the Large Mixing MSW is the solution of the
solar neutrino problem \cite{Hol,LMA_sun,Bahcall}.

In a reactor, anti neutrinos are released by radioactive isotope
fission; the total neutrino spectrum is a rather well understood function of
the thermal power $W$, the amount of thermal power $w_{i}$ emitted during
the fission of a given nucleus, and the isotopic composition of the reactor
fuel $f_{i}$,
\begin{equation}
S(E_{\nu })=\frac{W}{\sum f_{i}w_{i}}\sum f_{i}\left( \frac{dN}{dE_{\nu }}%
\right) _{i}
\end{equation}
The index $i$ of $f_{i}$ stands for 4 isotopes such are $^{235}U$,
 $^{238}U$,  $^{239}Pu$ and $^{241}Pu.$ The (dN/dE) is the
energy spectrum of the fissionable isotope, it can be parameterized by the
following expression\cite{nu_spectrum} when $E_{\nu }\geq $ $2MeV$:
\begin{equation}
\frac{dN_{\nu }}{dE_{\nu }}=e^{a_{0}+a_{1}E_{\nu }+a_{2}E_{\nu }^{2}}
\end{equation}%
the coefficients $a_{i}$ depend on the nature of the fissionable isotope.
KamLAND is a scintillator detector, where electronic anti neutrinos are
detected by free protons via inverse $\beta -$decay reaction\cite
{nu_spectrum},
\begin{equation}
\overline{\nu }_{e}+p\rightarrow e^{+}+n
\label{inverse-decay}
\end{equation}%
in the limit of infinite nucleon mass, the cross section of this reaction is
given by $\sigma (E_{\nu })=kE_{e^{+}}P_{e^{+}}$, where $%
E_{e^{+}},~P_{e^{+}} $ are the positron energy and momentum respectively and
$k$ can be taken as $k=9.55\times 10^{-44}~cm^{2}/MeV^{2}$. The anti neutrino
events are characterized by the positron annihilation signal and the delayed
neutron capture sign \cite{KamLAND}.

From the reactor to the detector, massive neutrinos oscillate on the way and
change their flavor composition to a certain extent. The anti neutrino, $
\overline{\nu }_{e}$, can oscillate to other flavors via $\Delta m_{12}^{2}$
and $\Delta m_{13}^{2}$. For KamLAND experiment, the distance is in the
range of hundred kilometers. Due to the tiny value of $\theta _{13}$ and the
value of $\Delta m_{13}^{2}$, contribution to oscillation from the
biggest neutrino mass scale $m_{3}$ gives a small averaged effect thus we
can reduce it to a two-flavor \ neutrino analysis:
\begin{equation}
P(E_{\nu },L,\theta _{12},\Delta m_{12}^{2})=1-\sin ^{2}(2\theta _{12})\sin
^{2}\left( \frac{1.27\Delta m_{12}^{2}(\mathtt{eV}^{2})L(\mathtt{m})}{E_{\nu
}(\mathtt{MeV})}\right)
\end{equation}

\begin{figure}[t]
\mbox{\epsfig{figure=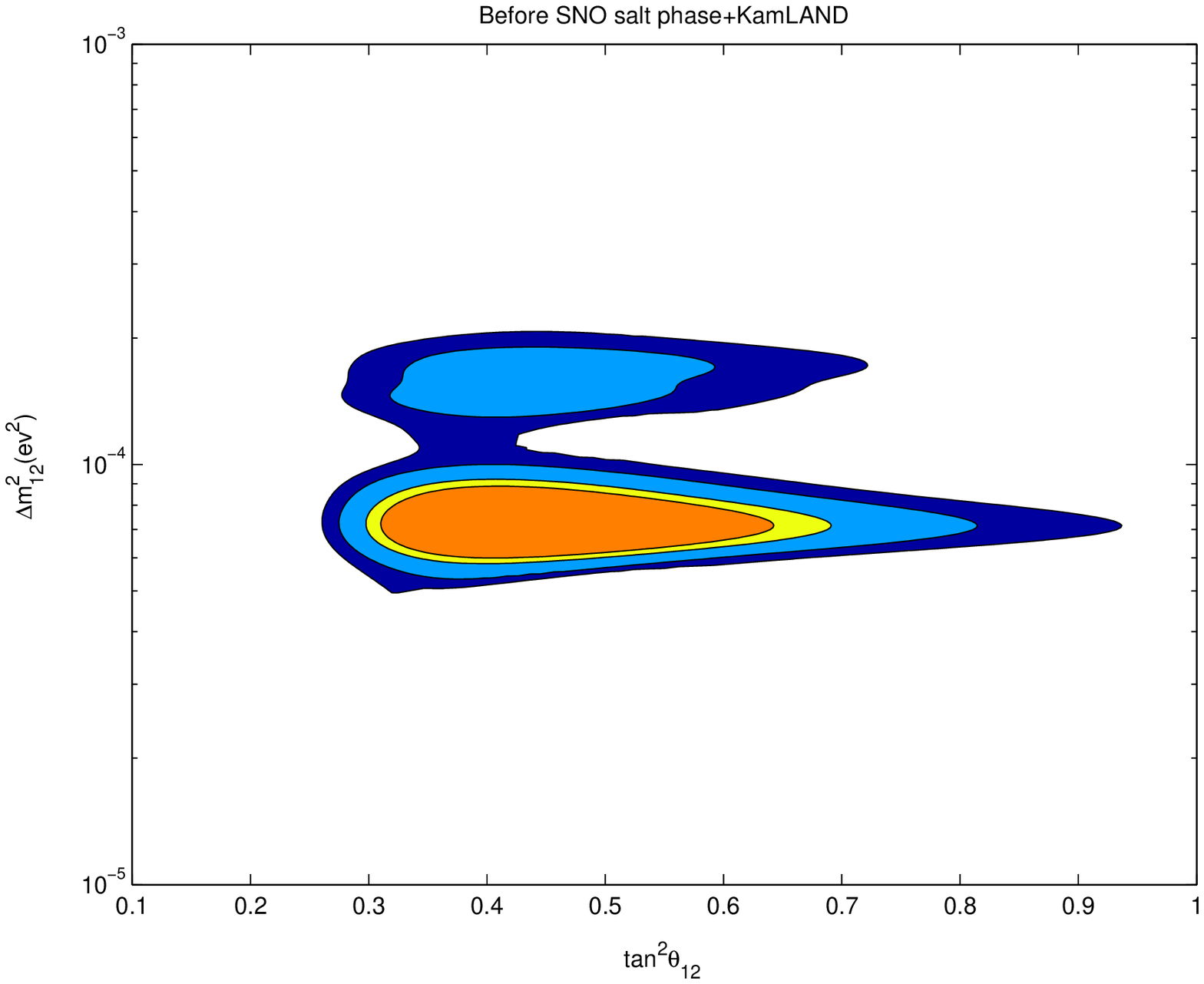,height=6cm}} \mbox{%
\epsfig{figure=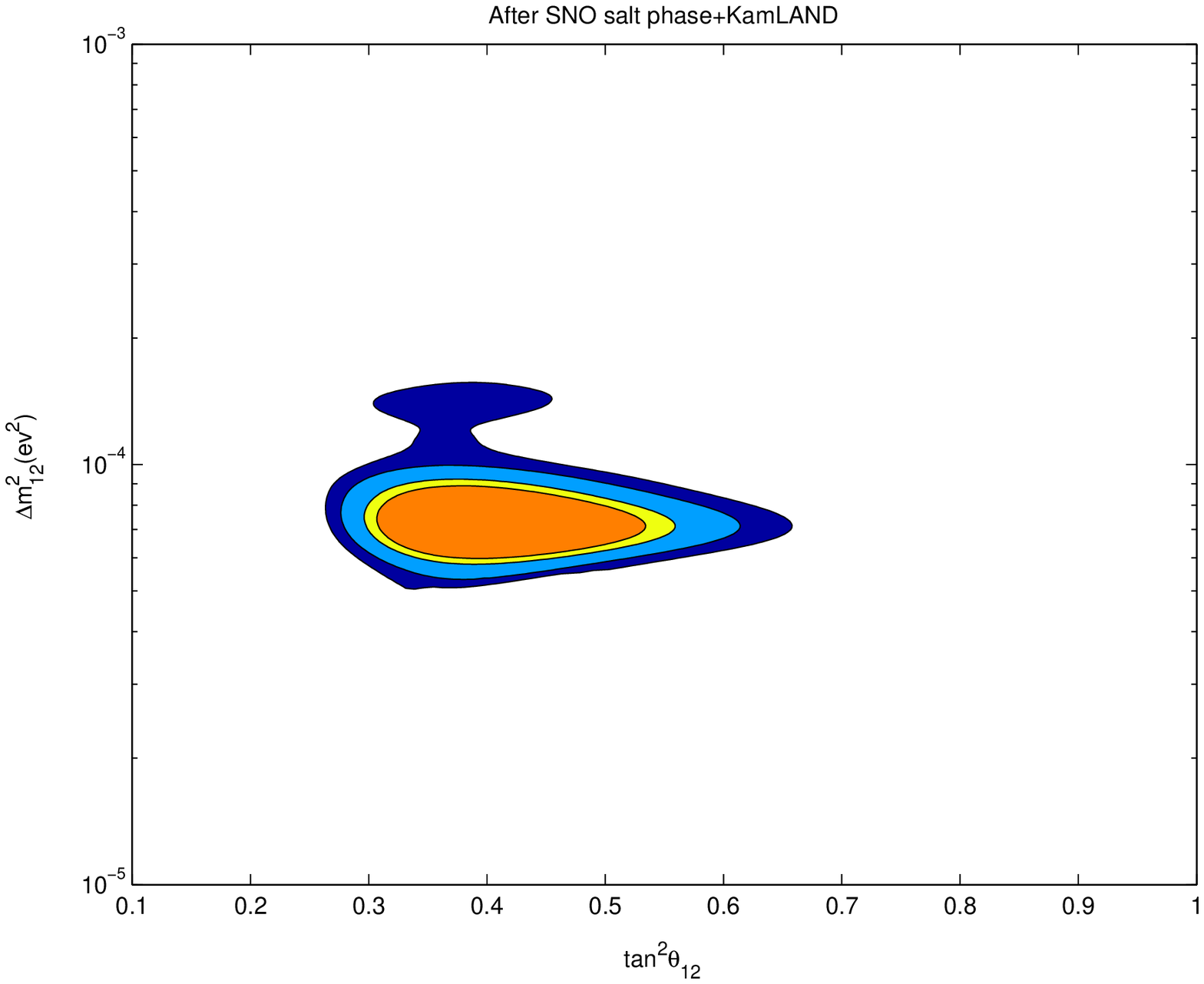,height=6cm}}
\caption{
  The combined LMA region of solar and KamLAND data before
 conference Nu$2004$. The difference between left and right shows
  improvement from SNO's salt phase result.
The confidence
  levels are the same as what in fig. \ref{fig:SNO}. }
\label{fig:BKamL}
\end{figure}

The combined results of solar and reactor neutrino experimental data before "Neutrino 2004" are shown in figs. \ref{fig:BKamL}. 

\begin{figure}[t]
\mbox{\epsfig{figure=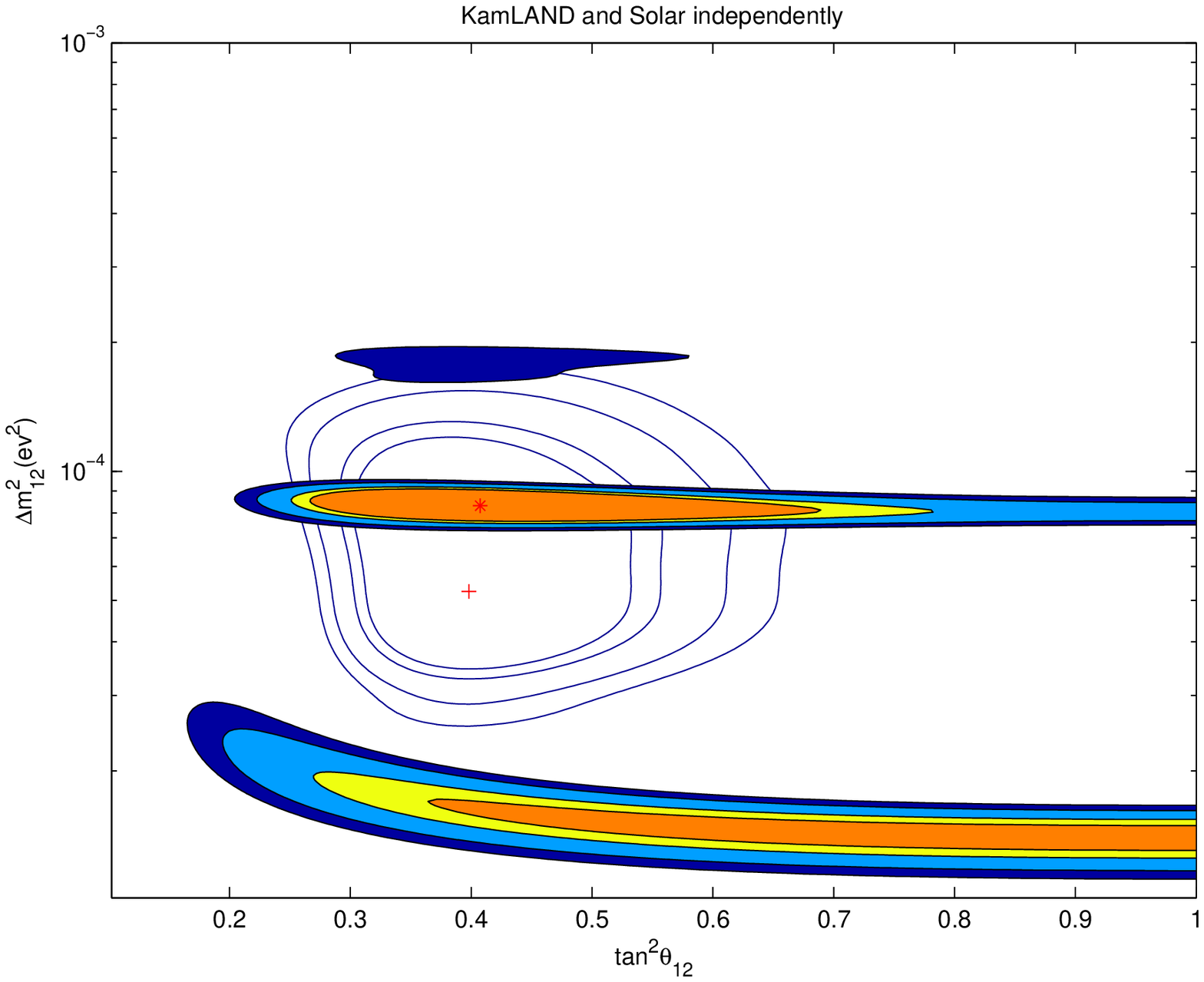,height=6cm}} \mbox{%
\epsfig{figure=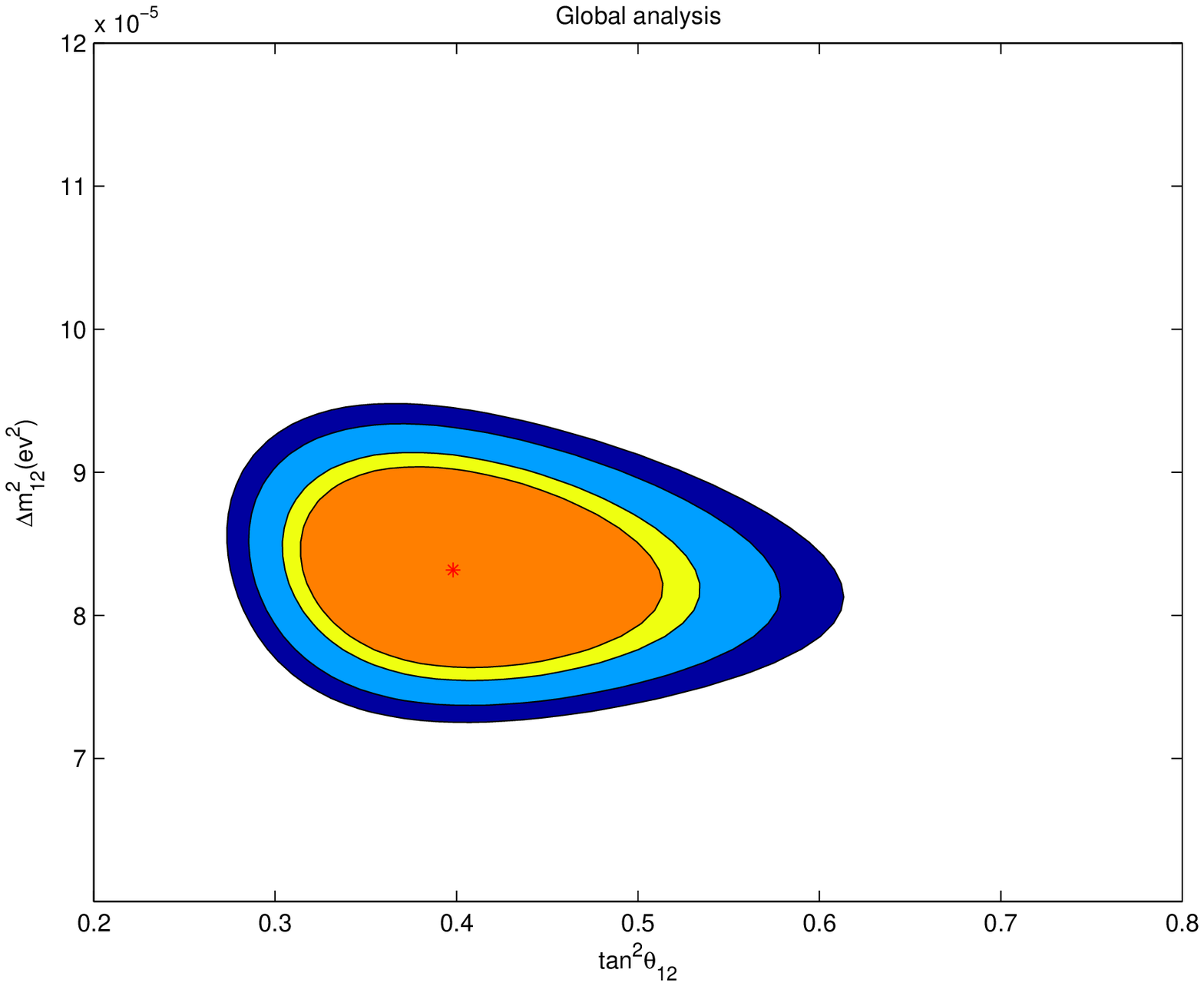,height=6cm}}
\caption{Left is the new KamLAND allowed regions for 766.3 Ty data,
  together with the LMA solution contours (not filled in) by solar
  neutrinos themselves. Right is
  the combined LMA region of solar and KamLAND data. The confidence
  levels are the same as what in fig. \ref{fig:SNO}. }
\label{fig:KamL}
\end{figure}

Using the newest 766.3 Ton-Year of data published by KamLAND collaboration
at conference $Neutrino$ 2004, we find the $\tan ^{2}\theta
_{12}-\Delta m_{12}^{2}$ allowed region with the energy spectrum in 13 bins.
The result is shown in fig. \ref{fig:KamL} -left. The combined chi
square is then $\chi^{2}=\chi _{\otimes }^{2}+\chi _{KL}^{2}$ with
result shown in fig. \ref{fig:KamL} -right;
which confirms the Large Mixing Angle MSW solution to the solar neutrino
problem. The best fit point we get is
% \begin{equation}
$$\tan^2\theta=0.4;   ~~~~~\Delta
m^2_{21}=8.3 \times 10^{-5}eV^2, $$
% \end{equation}
which is consistent with the previous papers \cite{analy}; The day-night effect \cite{daynight} for solar neutrinos
is then expected to be about two to four percent.

% and the ever happened solar neutrino energy spectrum
% puzzle is as predicted in paper \cite{solarspectrum}.

Many other possible candidates for the solution of solar neutrino problem are
excluded; sterile neutrinos are no more supported in this long-lasting
puzzle. However, the most reliable candidate, the LMA MSW solution is
becoming precise and more confident in terms of a much larger
confidence level and much smaller allowed region. Thus we conclude
that the solar neutrino problem is solved by large mixing MSW
solution, with parameters shown in fig. \ref{fig:KamL}.

  In conclusion, using the new data of KamLAND experiment we conclude
  that the solar neutrino solution is in
the large mixing MSW adiabatic allowed region, corresponding to
($\tan^2\theta$, $\Delta m^2$) $\sim$ ($0.27 - 0.62$,
$7.2 \times 10^{-5} - 9.5 \times 10^{-5} eV^{2}$), at $4\sigma$ level.
Which is an amazing small region in the parameter space with such a
high confidence level. Our result with best fit point at ($\tan^2\theta$,
  $\Delta m^2$) $\sim$ ($0.4$, $8.3 \times 10^{-5}
  eV^{2}$) is in good agreement with previous
  studies. Predictions of the MSW adiabatic
  conversion solution of the neutrino energy spectrum, the day-night effect,
  the seasonal variation, the ratio of CC and NC current events are
  all consistent with the solar
  data. At present, none signal indicates other solutions rather than the MSW
  adiabatic solution, together with the standard solar model.

\noindent{\large \textbf{Acknowledgments:}} One of the authors,
Q.Y.L., would like to thank A. Yu. Smirnov for reading of the
paper and useful suggestions, and the Abdus Salam International
Center for Theoretical Physics for hospitality when the paper was
written. This work is supported in part by the National Nature
Science Foundation of China.


\newpage

\begin{thebibliography}{99}


\bibitem{CHLOR} K. Lande, Talk given at the Neutrino '96 Int. Conference, June 13 - 19, 1996, Helsinki, Finland (to be published in the Proceedings of the Conference); see also: R. Davis, Prog. Part. Nucl. Phys. {\bf 32}, 13 (1994); B. T. Cleveland {\it et al.}, Nucl. Phys. B (Proc. Suppl.) {\bf 38}, 47 (1995).

\bibitem{SAGE} V. Gavrin et al. (SAGE Collaboration), Talk given at the Neutrino '96 Int. Conference, June 13 - 19, 1996, Helsinki, Finland (to be published in the Proceedings of the Conference); see also: J. N. Abdurashitov {\it et al.}, Phys. Lett. B {\bf 328}, 234 (1994).

\bibitem{GALLEX} P. Anselmann et al. (GALLEX Collaboration), Phys. Lett. B {\bf 357}, 237 (1995) (see also: ibid. B {\bf 327}, 377 (1994)).

\bibitem{Kamiokande}{K.S. Hirata, et al., Phys. Rev. D {\bf{44}}, 2241 (1991); K.S. Hirata, et al., Phys. Rev. Lett. {\bf{66}}, 9 (1991); Y. Fukuda et al., Phys. Rev. Lett. {\bf{77}}, (1996) 1683.  }

\bibitem{SNO} Q.R. Ahmad et al. [SNO Collaboration],
  Phys. Rev. Lett. 87, 071301 (2001);  Phys. Rev. Lett. 89, 011301 (2002).

\bibitem{KamLAND} K. Eguchi $et~al.$, Phys. Rev. Lett. 90, 021802 (2003).

\bibitem{Liu1} Q.Y. Liu and S.T. Petcov, Phys.Rev.D56: 7392 (1997);
Q.Y. Liu, proceedings of 4th International Solar Neutrino Conference,
Heidelberg, Germany, (1997) hep-ph/9708308.


\bibitem{Parke} Stephen J. Parke, Phys.Rev.Lett57, 1275 (1986).


\bibitem{Mont} E.Lisi, D.Montanino, Phys.Rev.D56, 1792 (1997).


\bibitem{Hol} P.C.de Holanda A.Yu.Smirnov Phys.Rev.D66, 113005 (2002)
  hepph/0205241.


\bibitem{BP} J.N.Bahcall, M.H.Pinsonneault, Phys. Rev. Lett. 92, 121301 (2004); J.N.Bahcall, M.H.Pinsonneault, S.Basu,Astrophus.J.555, 990 (2001).

\bibitem{LMA_sun} P.C. de Holanda and A. Yu.
Smirnov, J. Cosmol. Astropart. Phys. 02, 001 (2003).


\bibitem{Bahcall} J. N. Bahcall, M. C. Gonzalez-Garcia and C. Pe$\widetilde{%
\mathtt{n}}$a-Garay, High Energy Phys. 02, 009 (2003).


\bibitem{nu_spectrum} P. Vogel and J. Engel, Phys. Rev. D 39, 3378 (1989); H.
Murayama and A. Pierce, Phys.Rev. D65, 013012 (2002).

\bibitem{analy} [KamLAND Collaboration], hep-ex/0406035; J.N. Bahcall,
 M.C. Gonzale-Garcia, C. Pena-Garay, hep-ph/0406294; Abhijit Bandyopadhyay, Sandhya Choubey,
 Srubabati Goswami, S.T. Petcov, D.P. Roy, hep-ph/0406328.

\bibitem{daynight} Q.Y. Liu, M. Maris and S.T. Petcov,
Phys.Rev.D56, 5991 (1997); A. Dighe, Q.Y. Liu and A. Yu. Smirnov,
hep-ph/9903329.




%%%%%%%%%%%%%%%%%%%%%%%%%%%%%%%%%%%%%%%%%%%%%%%%%%%%%%%%%%%%%%%%%%%%%%%%%%%%%%%






% \bibitem{solarspectrum} Q.Y. Liu, proceedings of 17th International
% Workshop on "Weak Interactions and Neutrinos WIN '99", Cape Town, South Africa,
% 1999, hep-ph/9906553.


% \bibitem{MSW} S.P. Mikheyev, A.Yu. Smirnov, Sov. J. Nucl. Phys. 6, 913 (1985); L. Wolfenstein, Phys. Rev. D17, 2369 (1978);
%  S.P. Mikheyev and A.Yu. Smirnov, Yad. Fiz., {\bf 42},
%   1441 (1985); Nuovo Cim {\bf 9C}, 17 (1986).

% \bibitem{SK_atm} Y. Fukuda $et~al.$, Phys. Rev. Lett. 82, 2644 (1999); T.
% Toshito $et~al.$, hep-ex/0105023; S. Fukuda $et~al.$, Phys. Rev.
% Lett 86, 5651 (2001).


\end{thebibliography}
\end{document}